\documentclass[prd,preprint,tightenlines,floatfix,preprintnumbers,nofootinbib,eqsecnum]{revtex4}

\pdfoutput=1

 \usepackage[dvips,final]{graphicx}
  \usepackage{amssymb}
   \usepackage{amsmath}
    \usepackage{amsfonts}
     \usepackage{epsfig}
      \usepackage{bm} 
      \usepackage[dvipsnames,usenames]{color}
      \usepackage{comment}
      \usepackage{float}
      
      \usepackage[utf8]{inputenc}

\begin{document}

\title{Valence and sea parton correlations in double parton scattering from data}

\author{Edgar Huayra}
\email{yuberth022@gmail.com}

\author{João Vitor C. Lovato}
\email{joaovitorcl1000@gmail.com}

\author{Emmanuel G. de Oliveira}
\email{emmanuel.de.oliveira@ufsc.br}

\affiliation{
\\
{$^1$\sl Departamento de F\'isica, CFM, Universidade Federal 
de Santa Catarina, C.P. 476, CEP 88.040-900 Florian\'opolis, 
Santa Catarina, Brazil
}}

\begin{abstract}
\vspace{0.5cm}
The effective cross section of double parton scattering in proton collisions has been measured by many experiments with rather different results. 
Motivated by this fact, we assumed that the parton correlations in the transverse plane are different whether we have valence or sea partons.  
With this simple approach, we were able to fit the available data and found that sea parton pairs are more correlated in the transverse plane than valence--sea parton pairs.


\end{abstract}


\maketitle

\section{Introduction}
\label{Sect:intro}

The hard scattering of two hadrons producing a final state in the collinear factorization approach is described with success as a convolution of the parton cross section and parton distribution functions (PDFs). When there are two partons interacting, one from each hadron, the process is called single parton scattering (SPS). As energy increases, the number of incident partons increase, and there is a  greater probability of two or more hard scatterings occurring in the same inelastic event. This is known as multiple parton interactions (MPI)~\cite{Paver:1982yp, Diehl:2011yj}.

The particular case of MPI involving two partons from each hadron is called double parton scattering (DPS) \cite{Manohar:2012pe, Blok:2010ge, Blok:2011bu, Blok:2013bpa, Diehl:2014doa, Lewandowska:2014wga, PhysRevD.85.114009, 1987ZPhyC..34..163A, Kasemets:2013nma, Gaunt:2012tfk}. 
This important process in high-energy physics brings information about new QCD objects, namely, the double parton distribution functions (dPDFs), widely explored in the literature \cite{Chang:2012nw, Blok:2016lmd}. 
There are some models that describe the most significant features of dPDFs, but the availability of data is somewhat limited considering that dPDFs depend on two momentum fractions $x_1$ and $x_2$ \cite{Gaunt:2009re}. 

In the simplified scenario of neglecting all correlations among partons, the inclusive double parton scattering cross section for the two parton processes $A$ and $B$ can be described as the incoherent superposition of two single hard scattering processes:
\begin{eqnarray}
\sigma^{\text{DPS}} = \frac{m}{2} \frac{\sigma^{\text{SPS}}(A)\sigma^{\text{SPS}}(B)}{\sigma_{\text{eff}}}.
\label{pocketformula}
\end{eqnarray}
The $\sigma^{\text{SPS}}$ are the SPS cross sections for the final states $A$ and $B$, the symmetry factor $m$ is 1 if $A$ and $B$ are indistinguishable and $m=2$ otherwise. To complete this pocket formula, there is the effective cross section, $\sigma_{\text{eff}}$, that encodes all information about the transverse hadron structure. This simple hypothesis works quite well for the proton and is supported by the experimental evidence \cite{Leontsinis:2022cyi, ATLAS:2016ydt, Lansberg:2019adr, CMS:2019jcb, CMS:2021lxi, Gueta:2014rsi, CMS:2013huw, CMS:2015wcf, ATLAS:2016rnd, CDF:1993sbj, D0:2014owy, CDF:1997yfa, D0:2015rpo, CMS:2022pio, PhysRevD.90.111101, Lansberg:2017chq, LHCb:2016wuo, Shao:2016wor, Alitti1991ASO, Lansberg:2016muq}, where each result can be described using a single parameter (namely, $\sigma_{\text{eff}}$). However, the data suggests that this parameter is not universal, and a wide range of its values is found.

The starting point of our work is based on the idea that the average transverse distance between a pair of valence quarks in the hadron differs from that between a pair of gluons or sea quarks as well as the one between a valence and a sea parton. This may be a cause of the discrepancy found in the measurements of the effective cross section, because different final states will weight differently the sea and valence parton contributions. This actually has been explored previously using a hadron model to take into account the transverse distribution of  gluons and quarks \cite{DelFabbro:2000ds}. 

Our paper is simpler in that no model for the hadron is assumed, but we investigate valence and sea parton correlations by supposing that the $\sigma_{\text{eff}}$ depends on the participating partons. In the following, we shall quantify the valence and sea contributions for the DPS through a global fit to the effective cross sections of (anti)proton collisions measured in several processes with different final states by various collaborations. As we expected, the fit quality is not bad, showing that parton--kind correlations indeed are relevant.

The following is organized into three sections. First, in Section II, we provide a brief explanation of the theoretical framework used in DPS and of the idea used to describe the effective cross section considering the nontrivial parton correlations within the hadron. In Section III, we present the method employed to fit the experimental data and obtain values of $\sigma_{\text{eff}}$ depending on final states. Moreover, we show and discuss our results. Finally, in Section IV, we conclude our paper with a summary of our main findings.

\section{Theoretical framework
}
\label{Sect:formalism}

We first discuss formalism on which the following analysis is based. To work out the $\sigma_{\text{eff}}$ in the correlated case, one writes the inclusive double parton scattering cross section, for the hadrons $h$, $h'$ and final states $A$ and $B$ as
\begin{eqnarray}
\sigma^{\text{DPS}} &=& \frac{m}{2} \sum_{i j; k' l'} \int  dx_1 dx_2 dx_1' dx_2' d^2 r \ \Gamma_{i j}^h(x_1, x_2, \mathbf{r}) \hat{\sigma}_{i k'}^A (x_1, x_1') \hat{\sigma}_{j l'}^B (x_2, x_2') \Gamma_{k' l'}^{h'}(x_1', x_2', \mathbf{r}). \nonumber \\
& &
\end{eqnarray}
In this expression, $x_1$ and $x_2$ are the longitudinal momentum fractions of the partons from hadron $h$, while $x_1'$ and $x_2'$ are the longitudinal momentum fractions of the partons from hadron $h'$; the two-dimensional transverse distance between $ij$ or $k'l'$ partons is $\mathbf{r}$.
The sum $\sum_{i j; k' l'}$ is over all parton species and there is the convolution of double parton distribution functions $\Gamma_{i j}^h$ and $\Gamma^{h'}_{k' l'}$ and parton cross sections $\hat{\sigma}_{i k'}^{A,B}$. 
The factorization scale dependence of the dPDFs and of the parton cross sections was hidden to not overload the notation.

By neglecting longitudinal--transverse correlations, the double parton distribution $\Gamma_{i j}^h$ satisfy the factorized ansatz:
\begin{eqnarray}
    \Gamma_{i j}^h(x_1, x_2, \mathbf{r}) = f_{i/h}(x_1) f_{j/h}(x_2)\ F_{ij}(\mathbf{r}),
\end{eqnarray}
where $f_{i,j/h}$ are the single PDFs and $F_{ij}$ is the function which contains the transverse distribution information entering cross sections. If $F_{ij}$ do not depend on $i$ and $j$ one obtains Eq. \ref{pocketformula} and the effective cross section $\sigma_{\text{eff}}$ is universal and given by:
\begin{eqnarray}
     \frac{1}{\sigma_\text{eff}} = \int d^2 r\ F^2(\mathbf{r}).
\end{eqnarray}

The main proposition of this paper is that the functions $F_{ij}$ do depend whether we have valence or sea partons and then the DPS cross section is written as
\begin{eqnarray}
    \sigma^{\text{DPS}} = \frac{m}{2} \sum_{ij; k'l'} \Theta^{ij}_{k' l'} \sigma_{ik'}(A) \sigma_{jl'}(B).
    \label{factorizedcrosssection}
\end{eqnarray}
The hadron cross sections for two partons of species $i$, $j$ and $k', l'$ to undergo the hard interactions $A$ and $B$ respectively are given by
\begin{subequations}
\begin{eqnarray}
    \sigma_{ik'}(A) &=& \int dx_1 dx_1' \ \hat{\sigma}_{i k'}^A (x_1, x_1') f_{i/ h}(x_1) f_{k'/ h'}(x_1'), \\
    \sigma_{jl'}(B) &=& \int dx_2 dx_2' \ \hat{\sigma}_{j l'}^B (x_2, x_2') f_{j/ h}(x_2) f_{l'/h'}(x_2'),
\end{eqnarray}
\end{subequations}
and the total SPS cross section is
\begin{eqnarray}
    \sigma(A) = \sum_{ik'} \sigma_{ik'}(A). 
    \label{totalcrosssection}
\end{eqnarray}
The functions $\Theta^{ij}_{k' l'}$ are geometrical coefficients with dimension of the inverse of a cross section that contain information about parton correlations depending on the initial partons of the DPS processes and are written as
\begin{eqnarray}
     \Theta^{ij}_{k' l'}  
     = \frac{1}{\sigma^{ij}_{k'l',\text{eff}}} = \int d^2 r\ F_{ij}(\mathbf{r}) F_{k' l'}(\mathbf{r}).
     \label{scalefactor}
\end{eqnarray}

Using Eq.~\ref{pocketformula}, Eq.~\ref{factorizedcrosssection} and Eq.~\ref{totalcrosssection}, we define the double parton scattering effective cross section to be: 
\begin{eqnarray} \label{sigmaeff}
    \sigma_{\text{eff}}(AB)
    = \frac{m}2 \frac{\sigma(A)\sigma(B)}{\sigma^{\text{DPS}}}
    = \frac{\displaystyle \sum_{ik'} \sigma_{ik'}(A) \sum_{jl'} \sigma_{jl'}(B)}
    {\displaystyle  \sum_{ijk'l'} \sigma_{ik'}(A) \sigma_{jl'}(B)/\sigma^{ij}_{k'l',\text{eff}}}.
\end{eqnarray}
This $\sigma_{\text{eff}}(AB)$ depends on the $A$ and $B$ produced final states as each SPS cross sections weights differently the initial state partons. In this way, by knowing $\sigma_{\text{eff}}(AB)$ for a variety of final states, one can obtain the parton--kind dependent effective cross section $\sigma^{ij}_{k'l',\text{eff}}$ and thus extract some information about the transverse distributions of the hadrons.

\section{Effective cross section fit and results}
\label{Sect:prediction}

To calculate the effective cross section, we make the simplified assumption that the $\sigma_{\text{eff}}(AB)$ coefficients differentiate only between two kinds of partons: valence quarks and sea quarks and gluons. As a result, the indices $i$ and $k'$ can assume either the value of $v$ (which stands for ``valence'') or $s$ (``sea''). Therefore, there are $16$ possible coefficients, but by symmetry, several of them are equal. At the end, there are only $6$ independent coefficients, namely, $\sigma^{ss}_{ss,\text{eff}}$, $\sigma^{ss}_{sv,\text{eff}}$, $\sigma^{ss}_{vv,\text{eff}}$, $\sigma^{sv}_{sv,\text{eff}}$, $\sigma^{sv}_{vv,\text{eff}}$ and $\sigma^{vv}_{vv,\text{eff}}$. 

In order to determine the coefficients, we fit Eq.~\ref{sigmaeff} to the available data~\cite{Leontsinis:2022cyi, ATLAS:2016ydt, Lansberg:2019adr, CMS:2019jcb,CMS:2021lxi, Gueta:2014rsi, CMS:2013huw, CMS:2015wcf, ATLAS:2016rnd, CDF:1993sbj, D0:2014owy, CDF:1997yfa, D0:2015rpo, CMS:2022pio, PhysRevD.90.111101, Lansberg:2017chq, LHCb:2016wuo}. The data comprise various final states such as $J/\psi$, $\Upsilon$, jets, $Z$ and $W$ bosons and photons, from ATLAS, CMS, LHCb, CDF, and D0 experiments, thus we perform a global fit.  We use the library Minuit2 \cite{James2004MINUITUG} to minimize the reduced $\chi^2$ thus taking into account experimental uncertainties. 

The $\sigma_{ik'}(A,B)$ values at leading order (LO) are obtained with PYTHIA~8.3~\cite{Bierlich:2022pfr} and Madgraph~\cite{Alwall:2014hca}, both using the NNPDF23~\cite{NNPDF:2017mvq} as the parton distribution functions, taking into account the respective kinematic cuts of the experimental point corresponding to processes $A$ and $B$. We use the default parameters for Pythia, as well as its default PDF, because we consider that Pythia is trying to effectively describe as many high energy processes as possible, in the same spirit of our global fit. We expect that the results would not change much if we have used different Monte Carlo or PDF that tries to describe the global results using similar phenomenological models.

\begin{table}
	\def\arraystretch{1.4
	}
	\centering
	\begin{tabular}{|c|c|} \hline
		\,\,\, Effective cross section \,\,\,& \,\,\, Fit result (mb) \,\,\, \\  \hline \hline
		$\sigma^{ss}_{ss,\text{eff}}$ & $6.5 \pm 0.9$ \\ \hline
		$\sigma^{ss}_{sv,\text{eff}}$  &  $27 \pm 15$ \\ \hline 
	\end{tabular}
	\caption{Effective cross sections found in our fit with goodness  $\chi_{\text{dof}}^2 = 27.23/(18-2) = 1.70$. The notation $\sigma^{ij}_{k'l',\text{eff}}$ means that $i$ interacts with $k'$ and $j$ interacts with $l'$. The other variables  $\sigma^{sv}_{sv,\text{eff}}$, $\sigma^{ss}_{vv,\text{eff}}$,  $\sigma^{sv}_{vv,\text{eff}}$, and $\sigma^{vv}_{vv,\text{eff}}$ are fixed as the data is not sensitive to them. The result shows that sea partons are more correlated than other kinds.}
	\label{seavalence}
\end{table}

\begin{figure*}[t]
	\centering
	\includegraphics[width=\linewidth]{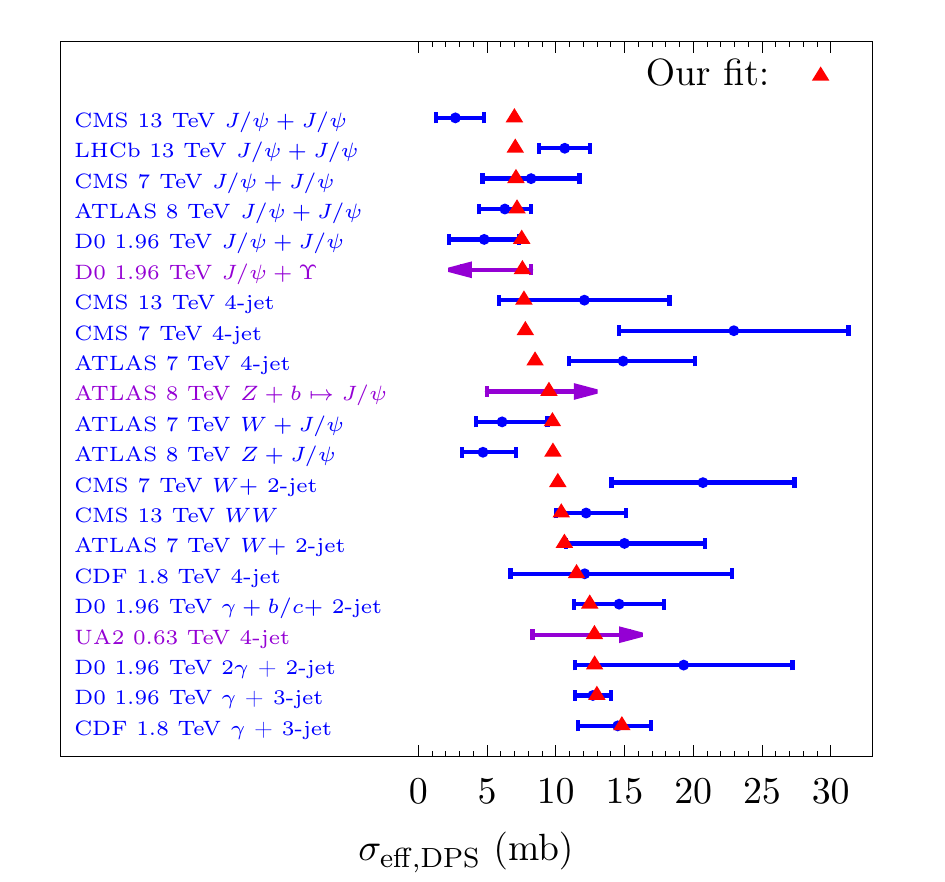}
	\caption{Double parton scattering effective cross section $\sigma_{\text{eff}}(AB)$ in several processes. The calculated and measured~\cite{Leontsinis:2022cyi, ATLAS:2016ydt, Lansberg:2019adr, CMS:2019jcb, CMS:2021lxi, Gueta:2014rsi, CMS:2013huw, CMS:2015wcf, ATLAS:2016rnd, CDF:1993sbj, D0:2014owy, D0:2015rpo, CMS:2022pio, PhysRevD.90.111101, Lansberg:2017chq, CDF:1997yfa, LHCb:2016wuo, Shao:2016wor, Alitti1991ASO, Lansberg:2016muq} values are compared. Only 8 of the 21 calculated values are outside the experimental error bars. Only blue datapoints are used in the fit, as purple arrows are upper or lower experimental limits.} 
	\label{results}
\end{figure*}

The minimization with the six $\sigma^{ij}_{k'l', \text{eff}}$ coefficients as free parameters does not fix the parameters $\sigma^{ss}_{vv,\text{eff}}$, $\sigma^{sv}_{sv,\text{eff}}$, $\sigma^{sv}_{vv,\text{eff}}$ and $\sigma^{vv}_{vv,\text{eff}}$, i.e., all the parameters with at least two valence quarks. This is due to the small values of SPS cross sections involving initial valence quarks. In other words, for most of the data, the strong ordering $\sigma_{ss}(A,B) \gg \sigma_{sv}(A,B) \gg \sigma_{vv}(A,B)$ is true. Thus, we do not take these 4 parameters as free and fix them to a constant value of 38\,mb that does not significantly change the results.

The results are shown in the Tab.~\ref{seavalence}, we find $\sigma^{ss}_{ss,\text{eff}} = 6.5 \pm 0.9$ mb and $\sigma^{ss}_{sv,\text{eff}} = 27  \pm 15$ mb. The goodness of the fit is $\chi_{\text{dof}}^2 = 27.23/(18-2) = 1.70$. The fact that $\sigma^{ss}_{sv,\text{eff}} > \sigma^{ss}_{ss,\text{eff}}$ indicates that sea partons are more correlated than the sea-valence combination and therefore, when participating in the DPS,  are closer in transverse space. Then, using the Eq.~\ref{factorizedcrosssection} we obtain the calculated effective cross sections for all processes.

In Fig.~\ref{results} we show the fitted experimental data as blue points with uncertainty bars and, as red triangles, the result from our fit. We also show three measured limits with purple arrows from ATLAS ($Z + b \rightarrow J/\psi$)~\cite{Lansberg:2016muq}, D0 ($J/\psi + \Upsilon$)~\cite{Shao:2016wor}, and UA2 ($4$-jet)~\cite{Alitti1991ASO}. 
The agreement between calculation and data is varied, but we remark that only 8 of the 21 calculated values are outside the experimental error bars.

We would like to comment that we expect that the valence-valence correlations are relatively weaker compared to the other parton kinds combinations, in fact, they maybe even be anticorrelated. For instance, the model used in Ref.~\cite{DelFabbro:2000ds} finds $\sigma^{sv}_{sv,\text{eff}} \approx \sigma^{sv}_{vv,\text{eff}} \approx \sigma^{vv}_{vv,\text{eff}} \approx 70$\,mb, showing some anticorrelation considering that one finds $38$\,mb by using only the proton radius in the uncorrelated case.

Additionally, we have tested the null hypothesis, in which the parton correlations do not depend on the parton-kind, i.e, the basic pocket formula of Eq.~\ref{pocketformula} holds. In this case, one gets $\sigma_\text{eff} = 9.8 \pm 0.6$\,mb with reduced chi squared given by $46.45/(18-1) = 2.73$. This gives a $p$-value of only $0.00015$ and as such the null hypothesis is rejected with confidence level of $3.8\sigma$. To reach $5 \sigma$ the experimental uncertainties would need to be multiplied by a factor of 0.87.

\section{Conclusions}
\label{Sect:conclusions}

In this work, we have studied the double parton distribution functions in the proton assuming that the distributions of valence and sea parton kinds could be different. In particular, we allowed different distributions in the transverse plane, perhaps due to correlations between a pair of partons from the same proton. This led to the usual pocket formula but with an effective cross section that is process dependent. We use this assumption to fit the double parton scattering effective cross section to data on several processes and from many experiments. The quality of the fit was not bad and the calculated values of the DPS effective cross sections are in good agreement with the experimental data, showing that allowing transverse correlations between parton populations is an important improvement in the description of DPS.

The parameters that could be fitted were $\sigma^{ss}_{ss,\text{eff}}$ and $\sigma^{ss}_{sv,\text{eff}}$, since for most measured DPS observables the valence contribution is rather small. We found that sea parton pairs are more correlated in the transverse plane than valence–sea parton pairs since $\sigma^{ss}_{ss,\text{eff}} < \sigma^{ss}_{sv,\text{eff}}$. More data on forward rapidity DPS observables and on final states such as $W$ boson and jets would help to determine valence--valence correlations. In particular, the $\sigma^{sv}_{vs,\text{eff}}=\sigma^{sv}_{sv,\text{eff}}$ effective cross section could benefit from a DPS measurement in which $A$ and $B$ final states have a large rapidity difference.  The effective cross sections corresponding to each parton distribution found in this way would give some quantitative idea about the transverse distance between respective parton kinds that could be included in current dPDF models.


\section*{Acknowledgments}

We would like to thank L.L.R.~Ferreira for fruitful discussions. This work was supported by FAPESC, INCT-FNA (464898/2014-5), and CNPq (Brazil) for EGdO, EH, and JVCL. This study was financed in part by the Coordenação de Aperfeiçoamento de Pessoal de Nível Superior -- Brasil (CAPES) -- Finance Code 001. 













\end{document}